# Dipole coupling of a hole double quantum dot in germanium hut wire to a microwave resonator

Gang Xu,[1,2,#] Yan Li,[1,2,#] Fei Gao,[3,4] Hai-Ou Li,[1,2,*] He Liu,[1,2] Ke Wang,[1,2] Gang Cao,[1,2] Ting Wang,[3,4] Jian-Jun Zhang,[3,4] Guang-Can Guo,[1,2] and Guo-Ping Guo[1,2,5*]

[1] *CAS Key Laboratory of Quantum Information, University of Science and Technology of China, Hefei, Anhui 230026, China*

[2] *CAS Center for Excellence and Synergetic Innovation Center in Quantum Information and Quantum Physics, University of Science and Technology of China, Hefei, Anhui 230026, China*

[3] *National Laboratory for Condensed Matter Physics and Institute of Physics, Chinese Academy of Sciences, Beijing 100190, China*

[4] *CAS Center for Excellence in Topological Quantum Computation, University of Chinese Academy of Sciences, Beijing 100190, China*

[5] *Origin Quantum Computing Company Limited, Hefei, Anhui 230026, China*

[#]these authors contributed equally to this work
[*]Emails: haiouli@ustc.edu.cn; gpguo@ustc.edu.cn

**Abstract:** The germanium (Ge) hut wire system has strong spin-orbit coupling, a long coherence time due to a very large heavy-light hole splitting, and the advantage of site-controlled large-scale hut wire positioning. These properties make the Ge hut wire a promising candidate for the realization of strong coupling of spin to superconducting resonators and scalability for multiple qubit coupling. We have coupled a reflection line resonator to a hole double quantum dot (DQD) formed in Ge hut wire. The amplitude and phase responses of the microwave resonator revealed that the charge stability diagrams of the DQD are in good agreement with those obtained from transport measurements. The DQD interdot tunneling rate is shown to be tunable from 6.2 GHz to 8.5 GHz, which demonstrates the ability to adjust the frequency detuning between the qubit and the resonator. Furthermore, we achieved a hole-resonator coupling strength of up to 15 MHz, with a charge qubit decoherence rate of 0.28 GHz. Meanwhile the hole spin-resonator coupling rate was estimated to be 3 MHz. These results suggest that holes of a DQD in a Ge hut wire are dipole coupled to microwave photons, potentially enabling tunable hole spin-photon interactions in Ge with an inherent spin-orbit coupling.

# 1. Introduction

The spin qubit formed by electrons trapped in silicon (Si) quantum dots has been recognized as a highly promising candidate for quantum computing because of its long coherence time and compatibility with mature semiconductor technology [1, 2]. Recent advances include fault-tolerant control fidelity for single-qubit gates [3, 4] and high fidelity for two-qubit gates [5-7] and strong spin-photon coupling [8-10]. These results are believed to enable the construction of a quantum computer based on spin qubits with photonic interconnections for long-range coupling to and hybrid integration with other quantum systems [11, 12]. However, to implement electron spin qubits in Si, an integrated component, such as a micromagnet [3, 5, 6] or a strip line [4, 7], has to be incorporated, which will complicate the fabrication process of such devices for spin control and spin-charge hybridization. Holes in Si and germanium (Ge) have strong spin–orbit interaction [13, 14], which can be implemented for fast and all-electrical control of spin qubit without any additional components. This has enabled the demonstration of single-qubit control in Si [15], single and two-qubit logic in Ge [16, 17], providing a scalable approach towards for quantum information processing.

As a group-IV element, Ge exists as a nuclear spin-0 isotopes, similar to Si, which has a weak hyperfine interaction [18, 19]. Therefore, the hole spin qubits in Ge possess a long coherence time [17]. The Ge/Si core/shell nanowire, a hole material grown by chemical vapor deposition, has been investigated for years [20-22]. Recently, a type of hole material named Ge hut wire grown by molecular beam epitaxy attracted growing attention [23-29]. Compared with the Ge/Si core/shell nanowire, the Ge hut wire is a different hole system because of its distinctive growth mechanism. First, heavy hole (HH) and light hole (LH) levels are sufficiently split [24] to suppress HH-LH mixing in Ge hut wire owing to their triangular cross-section with a height of approximately 2 nm above the wetting layer and fully strained lattice structure [23]. This splitting results in non-Ising type like hyperfine interactions in the Ge hut wire, which is beneficial for the spin coherence time [18, 19]. This phenomenon is in contrast with that in Ge/Si core/shell nanowires, where their cylindrical geometry induces the strong HH-LH mixing. Furthermore, due to the special growth mechanism called the Stranski-Krastanow growth mechanism [29], the Ge hut wire system allows site-controlled hut wire positioning on a large-scale patterned substrate by means of planar epitaxial growth [30], which presents an important advantage for the scalability of quantum computation. Previous studies on Ge hut wire have shown the transport measurements of quantum dots [24-26], single–shot reflectometry of the hole spin states [27] and the hole spin-orbit qubit [16]. Recently, we demonstrated the coupling of a Ge hut wire single quantum dot (SQD) to a microwave resonator [28]. Compared with the SQD system, the energy level splitting in the double quantum dot (DQD) can be tuned to an energy scale close to that of resonator photons. Therefore, to realize a tunable coherent spin-photon interaction, the architecture of a DQD dipole coupled to a resonator is preferable. Such hybrid devices have been implemented using a variety of materials, including Si [8, 9], GaAs [10, 31], carbon nanotubes [32, 33], graphene [34, 35], InAs nanowires [36, 37], InSb nanowires [38], and Ge/Si core/shell nanowires [39].

Here, we report a Ge hut wire hole DQD dipole coupled to a microwave resonator. The DQD charge stability diagrams extracted from measurements of the amplitude and phase of a microwave tone reflected from the resonator are in good agreement with those obtained from transport measurements. Characteristic parameters of the hybrid device, including the hole–resonator coupling strength, the charge qubit decoherence rate and the interdot tunneling rate, were extracted from the interdot charge transition line. Furthermore, we estimated the spin–resonator coupling strength and demonstrated the tunability of the DQD interdot tunnel coupling.

## 2. Experimental setup

The top panel of figure 1(a) shows the optical micrograph of the hybrid device (the resonator is connected to DQDs), and the bottom panel is a scanning electron microscopy (SEM) image of Ge hut wire DQD. The SEM image of the hybrid device is shown in figure A (d) of Appendix A. The resonator was fabricated by 200-nm-thick aluminum with the characteristic impedance ~50 Ω. Figure 1(b) shows a sketch map of the hybrid device, the hybrid architecture includes a half-wavelength (λ/2) reflection line resonator and four Ge hut wire DQDs. Only DQD2 was used in this study, and all the gates relevant to other DQDs were grounded. The resonator consists of two paralleled superconducting striplines, and each of them has two ends that can be connected to DQDs. The resonator is also connected to AC pads by finger capacitance for the input and output of the microwave. The electrical potentials on the two striplines create electromagnetic fields with opposite signs. In contrast with transmission line resonators based on coplanar waveguides (CPWs), no ground plane is used in our device. The resonator couples to qubits via a symmetric differential excitation, which potentially has a larger coupling strength and immunity to common-mode noise [40, 41]. The schematic of the measurement setup and equivalent circuit of the DQD are shown in figure 1(c). A continuous microwave signal was split into two differential components with opposite phases and then applied to the striplines to establish an electromagnetic field. Microwave reflectometry was performed using a network analyzer together with a primary cryogenic amplifier and a secondary room-temperature amplifier. The sample was anchored to the mixing chamber of a dilution refrigerator with a base temperature of 18 mK.

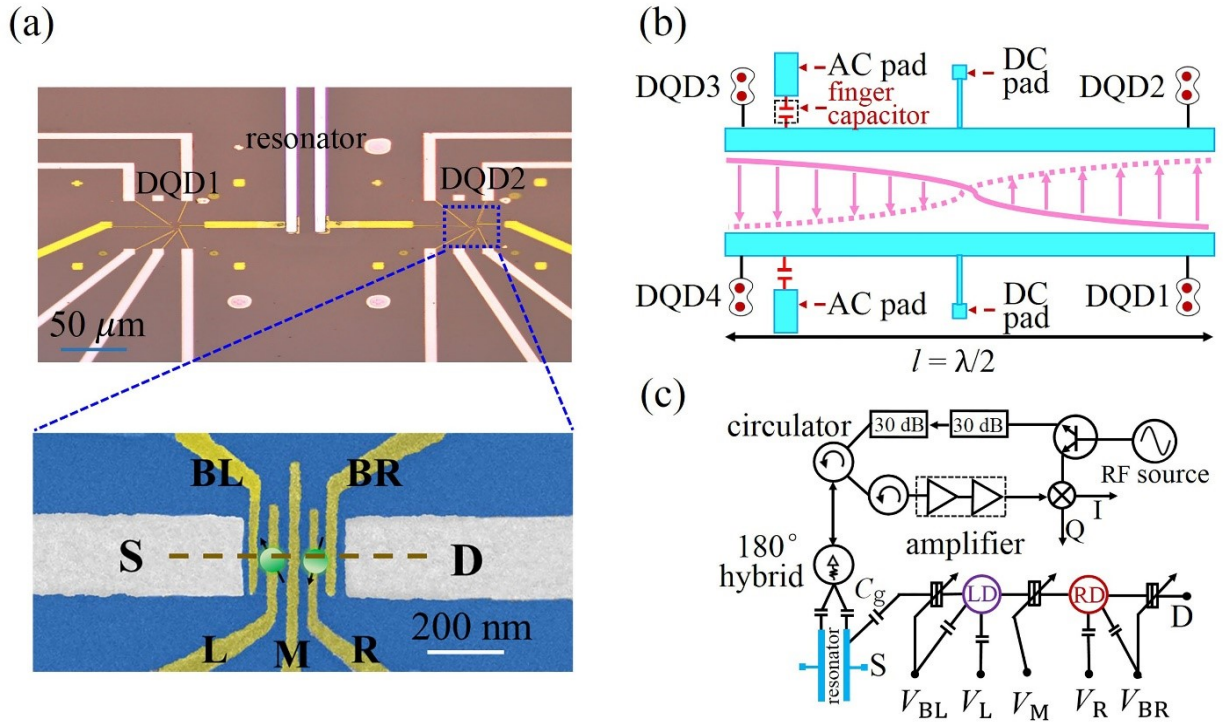

**Figure 1.** (a) Top panel: Optical micrograph of the hybrid device (the resonator connected to DQD). Bottom panel: False-colour scanning electron microscopy (SEM) image of a Ge hut wire DQD. The DQD is coupled to a microwave resonator through its lead (S). The left and right barrier gates (BL and BR), left and right plunger gates (L and R), and middle gate (M) are used to tune the chemical potential of the DQD. (b) Schematic of the device. The length of the half-wavelength ($\lambda/2$) resonator is $l$ =12.5 mm. It consists of two superconducting striplines, which are coupled to the microwave input and output ports via finger capacitors. The electrical potentials on the two striplines create steady electromagnetic fields with opposite signs. (c) Schematic of the measurement setup and equivalent circuit.

## 3. Resonator responses and charge stability diagrams of the DQD

We first characterized the microwave resonator by analyzing the amplitude and phase of the frequency spectrum of the reflected signal. The black line in figure 2 (a/b) represents a spectrum of the bare resonator, showing a Lorentzian line shape with a resonance frequency $f_0=\omega_0/2\pi = 6.038$ GHz. On the basis of the $\lambda/2$ open-circuit microstrip resonator model [40, 42], we extracted an internal loss rate, an external loss rate and a total loss rate ($\kappa_i/2\pi$, $\kappa_e/2\pi$, $\kappa/2\pi$) of approximately 2.6, 4.0, and 6.6 MHz, with a quality factor $Q = \omega_0/\kappa$ of 820. The power of the probe microwave signal reaching the input port of the resonator is $P_{in} \sim -110$ dBm throughout the experiment, corresponding to n ~40 photons in the bare resonator by rough estimation with n ~ $2P_{in}/\hbar f_0 \kappa$ [43].

We employed the microwave resonator to probe the properties of the Ge hut wire hole DQD. The probe frequency was fixed at the resonance frequency of the resonator. When the charge states of the DQD changed, the reflected amplitude and phase of the microwave signal changed correspondingly due to the dispersive coupling between the DQD and resonator. As shown in figures 2(a) and 2(b), both an amplitude shift $\Delta A$ and a phase shift $\Delta\phi$ are observed when the DQD is tuned from the Coulomb blockade regime [figure 2(c), blue rectangle] to the Coulomb resonance [figure 2(c), yellow cycle] regime.

During the measurement, the DC transport signal through the source and drain contacts was recorded by a multimeter after being passed through a low-noise preamplifier at a source-drain bias voltage of 0.1 mV. Figures 2(d) and 2(e) present the charge stability diagrams according to the amplitude and phase signals independent from the DC transport measurements in figure 2(c). These three diagrams show similar honeycomb structures, which indicate that the microwave signal can be used to detect the charge states of the Ge hut wire DQD. However, the amplitude and phase responses are found to behave differently from the transport measurements for the same honeycomb structure. From the DC transport measurements [figure 2(c)], the resonance to the right lead (indicated by the white dashed lines) is clearly visible. However, in the corresponding microwave measurements, as shown in figures 2(d) and 2(e), the resonance to the left lead is more pronounced. The DQD is connected to the microwave resonator through the left lead, which indicates a stronger capacitive coupling of the resonator to the left lead than to the right lead, thus the resonator is much less sensitive regarding tunneling events to the right lead [44].

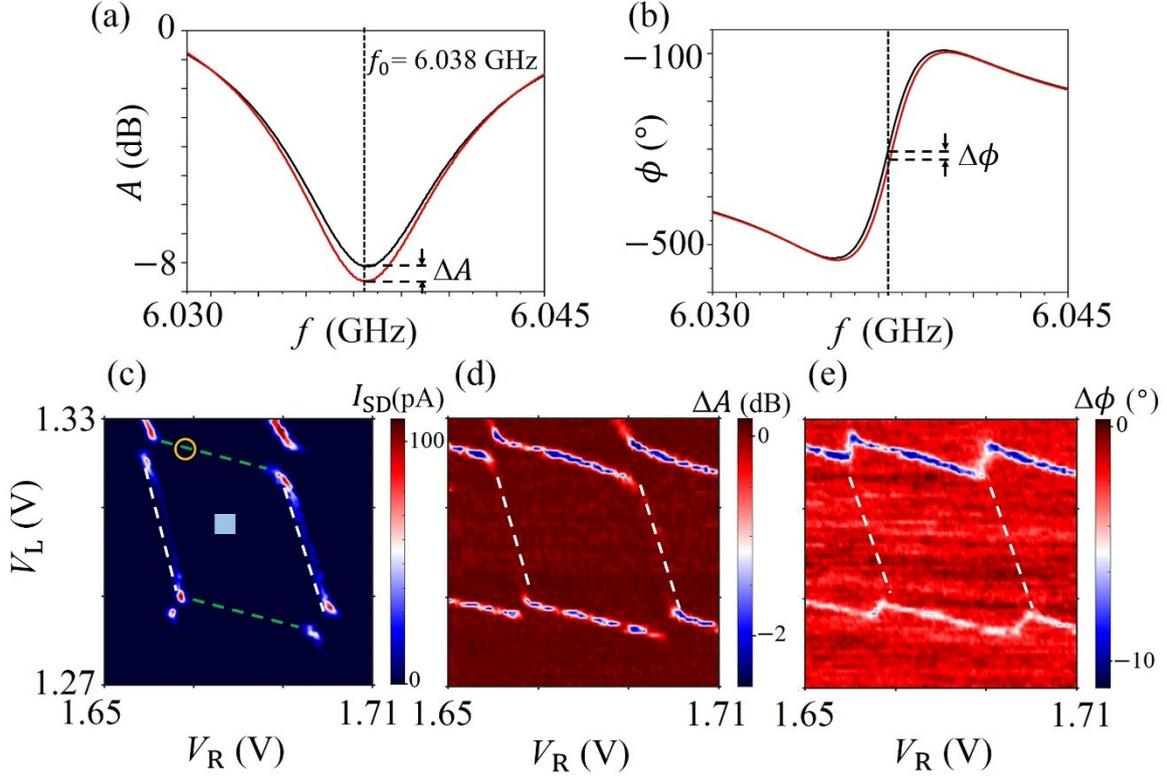

**Figure 2.** (a) Amplitude and (b) phase responses of the resonator as functions of the driving frequency. The black curves correspond to resonator responses when the DQD is in the Coulomb blockade regime. The red curves correspond to resonator responses when the DQD state is in the Coulomb resonance regime. Variations in the amplitude $\Delta A$ and phase $\Delta \phi$ can be seen clearly. (c) Transport signal of a stability diagram of DQD. Holes tunneling between the left (right) dot and reservoir (drain) is indicated by the green (white) dashed lines. (d) Amplitude and (e) phase responses of the resonator as functions of $V_L$ and $V_R$ over the same gate voltage range as in (c).

To study the interaction between a hole trapped in a DQD and the electric field of the resonator, we focus on the resonator responses near the (M, N+1) ↔ (M+1, N) interdot charge transition, where M and N are numbers of holes in left dot and right dot. The measurements were performed in many-hole regime (more than 10 holes in each dot). As show in figure 3(a), the DQD forms a two-level system with an energy splitting of $\Omega = \sqrt{(2t_C)^2 + \varepsilon^2}$, where $\varepsilon$ is the detuning. Interdot tunnel coupling hybridizes the charge states around $\varepsilon \sim 0$, resulting in a tunnel splitting of $2t_C$. To account for photon exchange between the microwave field and the DQD and investigate the interdot tunneling via the phase response, we employed the Jaynes-Cummings model [45] which has been used in a previous experimental study [36]. Figure 3(b) shows the reflected microwave phase response near the (M+1, N) ↔ (M, N+1) interdot charge transition. We analyzed the interdot tunneling along the detuning line [figure 3(b); green dashed arrow]. For this measurement, the phase shift depends on the resonance frequency $\omega_0$, the probe frequency $\omega$, the internal and external resonator dissipation rates $\kappa_i$ and $\kappa_e$, the DQD interdot tunneling rate $2t_C$, the hole-resonator coupling strength $g_C$, and the DQD susceptibility $\chi$. The coefficient of reflection is expressed as

$$S_{11}(\omega) = -\frac{j(\omega_0-\omega)+g_{\text{eff}}\chi+\frac{1}{2}(\kappa_i-\kappa_e)}{j(\omega_0-\omega)+g_{\text{eff}}\chi+\frac{1}{2}(\kappa_i+\kappa_e)}. \quad (1)$$

The phase and amplitude are extracted from $\phi = \arg(S_{11})$ and $A = |S_{11}|$. In equation (1), $\chi = \frac{g_{\text{eff}}}{j(\Omega-\omega)+\gamma}$ characterizes the susceptibility of the DQD. Here, $g_{\text{eff}} = g_C \frac{2t_C}{\Omega}$ is the effective coupling strength between the DQD and the resonator. The total decoherence rate of a charge qubit is $\gamma = \frac{1}{2}\gamma_1+\gamma_\emptyset$, with the energy relaxation rate $\gamma_1$ and dephasing rate $\gamma_\emptyset$. The measured phase shifts along the detuning line are plotted in figure 3(c), with the best-fitting curve obtained using equation (1). From the fitting results, we can determine that the interdot tunneling rate $2t_C/h$ is approximately 6.20 GHz. The hole-resonator coupling strength $g_C/2\pi$ is ~15 MHz and charge qubit decoherence rate $\gamma/2\pi$ is ~ 0.28 GHz. The charge qubit decoherence rate is an order of magnitude lower than that in the Ge/Si core/shell nanowire system ($\gamma/2\pi$~ 4-6 GHz [39]), comparable to previous reports on carbon nanotubes [32] and graphene systems [35], but still much larger than that recently reported on Si and GaAs systems (tens or few MHz) [46, 47]

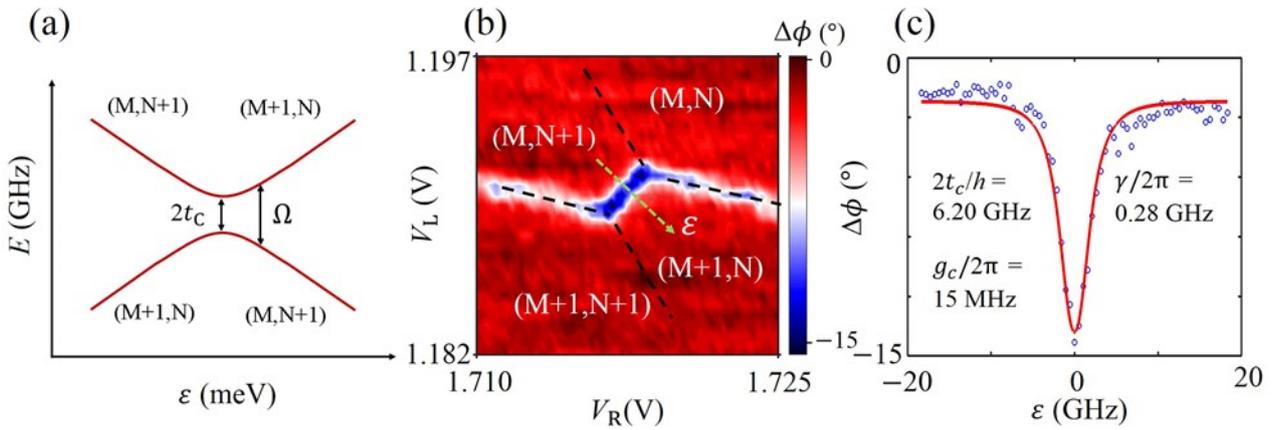

**Figure 3.** (a) Energy levels of the DQD obtained from $\Omega=\sqrt{(2t_C)^2+\varepsilon^2}$ as a function of the detuning $\varepsilon$; at $\varepsilon = 0$, the energy has the minimum value of $2t_C$. (b) Charge stability diagram obtained from the phase response near the (M+1, N) ↔ (M, N+1) interdot charge transition. The green dashed arrow marks the detuning line. (c) Fitted phase response according to equation (1) (red line) as a function of $\varepsilon$ obtained from measurements (blue dots) along the detuning line.

## 4. Estimation of the spin-resonator coupling rate

Although the electron spin states cannot be directly coupled to an electric field, the spin-orbit interaction enables electrical control of the spin by acting on the orbital component of the electron wave function. The spin-orbit interaction mixes spin and orbital degrees of freedom, resulting in spin states that have some orbital characteristics. Therefore, the use of spin-orbit interaction is a feasible way to achieve spin-photon coupling.

The spin-photon coupling strength of the DQD is then estimated as $g_S \approx 2g_C(\Delta E_0 E_Z/E_{qb}^2)(L/\lambda_{SO})\,\eta$, where $\eta = s/\sqrt{1-s^2}$ and $s = e^{-(L/l)^2}$ [39], $g_C$ is the hole-resonator coupling strength, $E_Z$ is the Zeeman splitting of the spin states, $\Delta E_0$ is the orbital level spacing, $l$ is the QD size and $\lambda_{SO}$ is the spin–orbit length that characterizes the strength of the spin-orbit interaction. To obtain the spin-photon coupling induced by the spin-orbit interaction, the photon energy should be close to the Zeeman splitting, which is assumed to be of the same order of magnitude as the energy of the lowest cavity mode [48], $E_Z = hf_0 \sim 25$ μeV, corresponding to an external magnetic field B of ~ 0.1 T and a *g* factor of ~ 4. The orbital level spacing $\Delta E_0$ is approximately 1 meV and the size of a Ge hut wire QD $l$ is~ 20nm as presented in a previous work [26]. Here, we choose $\lambda_{SO}\sim 40$ nm for HH states in Ge hut wires [26], the qubit energy $E_{qb} \approx 2t_C \approx 40$ μeV at $\varepsilon = 0$ and the half interdot distance $L \approx 40$ nm. Then the spin-photon coupling strength of DQD ($\varepsilon = 0$) is estimated to be $g_S/2\pi \sim 3$MHz. When the hole spin is trapped in a SQD ($\varepsilon \neq 0$), $g_S$ is expressed as $g_S \approx g_C(E_Z/\Delta E_0)(l/\lambda_{SO})$ [36, 48], then the spin-photon coupling strength of SQD is estimated to be $g_S/2\pi \sim 0.22$ MHz, which is comparable to the spin-resonator coupling rate $g_S/2\pi \sim 0.2$ MHz obtained in InAs nanowires [36]. By optimizing the resonator design and employing a high-impedance resonator [49, 50], the spin-resonator coupling rate $g_S$ can be enhanced up to one order of magnitude. Considering the long dephasing time in Ge hut wire [16], it reveals the possibility to achieve strong hole spin-photon coupling in future works.

## 5. The $V_M$-dependent interdot tunneling rate

We further studied the phase signal of the resonator in the same regime as shown in figure 3(b) but for different interdot tunneling rates, which were tuned by the middle gate voltage $V_M$. The phase responses near the same interdot charge transition line are plotted for several values of $V_M$ in figures 4(a)-4(d). From the reflection spectra, we observed a clear phase shift around the interdot charge transition line, and the line width was also broadened when $V_M$ increased. In the dispersive regime, the resonator frequency shift is $\sim g_{\text{eff}}^2/|\Omega - \omega_0|$[43]. Therefore, an increment of the interdot tunneling rate makes also the resonator-qubit frequency detuning higher.

From the evolution of phase spectra of the resonator for various middle gate voltages $V_M$ [figure 4(e)], we identify that the interdot tunneling rate $2t_C/h$ ranges from 6.2 to 8.5 GHz [figure 4(f), red points]. We did not obtain the lower value of $2t_C/h$, the saturation of $2t_C/h$ (~6 GHz at $V_M$ < 1.4 V in our system) is usually influenced by the high hole temperature $T_h$. However, since the $T_h$ here is estimated to be ~125 mK and it should not be high enough to influence the saturation of $2t_C/h$ according to the limit $k_B T_h \lesssim 2t_C$ [51]. A probable reason could be the traps and disorder occur near the working area that suppress the ability to tune $2t_C$. We obtained the decoherence rate $\gamma/2\pi$ that ranges from 0.28 to 2.32 GHz [figure 4(f), blue points]. The quite high charge qubit decoherence rate may attributed to two reasons. Firstly, by tuning the tunnel coupling a more positive $V_M$ is applied, leading to a large electric field which can untrap charges generating a higher noise level [52], where the strong charge noise has a negative effect on dephasing. Secondly, our measurements were performed at a relatively high RF power (the phase signal is difficult to measure at a lower RF power), a large drive power will also broaden the interdot transition line [53].

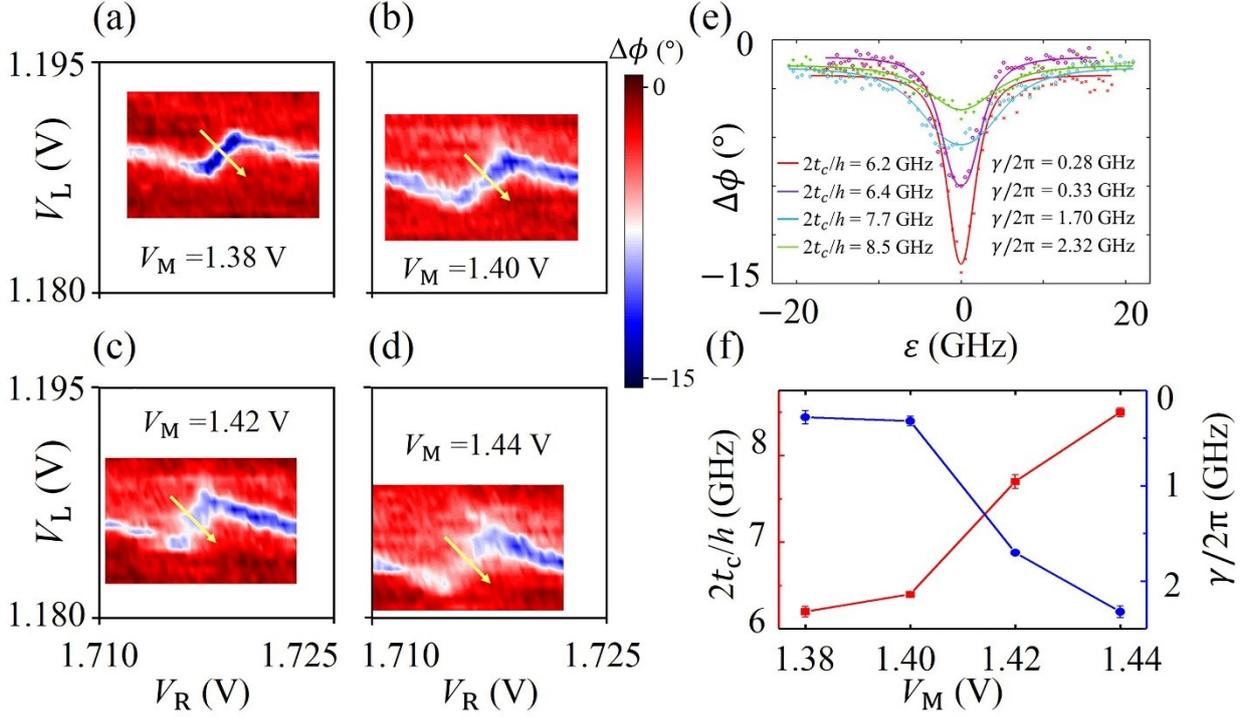

**Figure 4.** (a)-(d) Phase response near the interdot charge transition line for various middle gate voltages $V_M$ [see figure 1(a)]. As $V_M$ increases, the width of the interdot charge transition line is broadened, and the line becomes faint. (e) Phase response measured as a function of the DQD $\varepsilon$ [see yellow arrow in panels (a-d)]. The fitted parameters according to equation (1) are $[2t_C/h, g_C/2\pi, \gamma/2\pi]$ = [6.2 GHz, 15.0 MHz, 0.28 GHz] ($V_M$=1.38 V), $[2t_C/h, g_C/2\pi, \gamma/2\pi]$ = [6.4 GHz, 14.5 MHz, 0.33 GHz] ($V_M$=1.40 V), $[2t_C/h, g_C/2\pi, \gamma/2\pi]$ = [7.7 GHz, 16.6 MHz, 1.70 GHz] ($V_M$=1.42 V), $[2t_C/h, g_C/2\pi, \gamma/2\pi]$ = [8.5 GHz, 17.8 MHz, 2.32 GHz] ($V_M$=1.42 V). (f) Interdot tunnel rate $2t_C/h$ (read points) and decoherence rate $\gamma/2\pi$ (blue points) as a function of DQD middle gate voltages $V_M$.

## 6. Conclusions

In conclusion, we have demonstrated dipole coupling of a Ge hut wire hole DQD to a reflection line microwave resonator. The charge stability diagrams of the DQD obtained from the amplitude and phase of a microwave tone reflected from the resonator were demonstrated, and we obtained a tunable interdot tunneling rate ranging from 6.2 GHz to 8.5 GHz. Furthermore, the interdot charge transition line was analyzed by the Jaynes-Cummings model. A hole-resonator coupling strength $g_C/2\pi$ of up to 15 MHz was achieved in the experiment with an extracted charge decoherence rate $\gamma/2\pi$ of 0.28 GHz, and the spin-resonator coupling rate $g_s/2\pi$ was estimated to be 3 MHz. These results indicate that our architecture based on Ge hut wire could offer a way to probe the hole spin system in the microwave regime and reveals the potential to achieve strong hole spin-photon coupling in future works.

**Acknowledgements**

This work was supported by the National Key Research and Development Program of China (Grant No.2016YFA0301700), the National Natural Science Foundation of China (Grants No. 61674132, 11674300, 11625419 and 11574356), the Strategic Priority Research Program of the CAS (Grant Nos. XDB24030601 and XDB30000000), the Anhui initiative in Quantum information Technologies (Grants No. AHY080000) and this work was partially carried out at the USTC Center for Micro and Nanoscale Research and Fabrication.

# Appendix A. The growth method of germanium hut wire and more details about the device

Figures A (a) and (b) show the Ge hut wire [23] we study here. A 100-nm-thick Si buffer layer was first grown on an intrinsic Si substrate. Then 6.5 Å thick Ge was deposited on the Si buffer layer to form the Ge hut cluster structure at 540 °C. After an annealing process of 8 hours at 530 °C, these clusters transform into hut wires with lengths ranging from several hundred nanometers to approximately 1 μm. In the last step of the growth process, a 3.5-nm-thick Si capping layer was deposited at 330 °C to protect the Ge wire from oxidation. Figure A (c) shows a simplified three-dimensional schematic of a Ge hut wire DQD with five gates. The DC transport signal through the source/drain leads was measured by a multimeter after being passed through a preamplifier. The five top gate voltages can be changed to tune the DQD. Figure A (d) shows the patterning of the sample performed by electron beam lithography. The source and drain electrodes were metallized with a 30-nm-thick Pd layer after a short oxide removal step with buffered hydrofluoric acid. After a 30-nm-thick alumina dielectric layer was grown by atomic layer deposition, the superconducting resonator was patterned by optical lithography, 200-nm-thick Al was deposited, and the alumina around the source electrode was etched to allow a connection between the electrodes and the resonator. Finally, the top gate electrodes were subsequently fabricated with 3-nm-thick Ti and 25-nm-thick Pd layers.

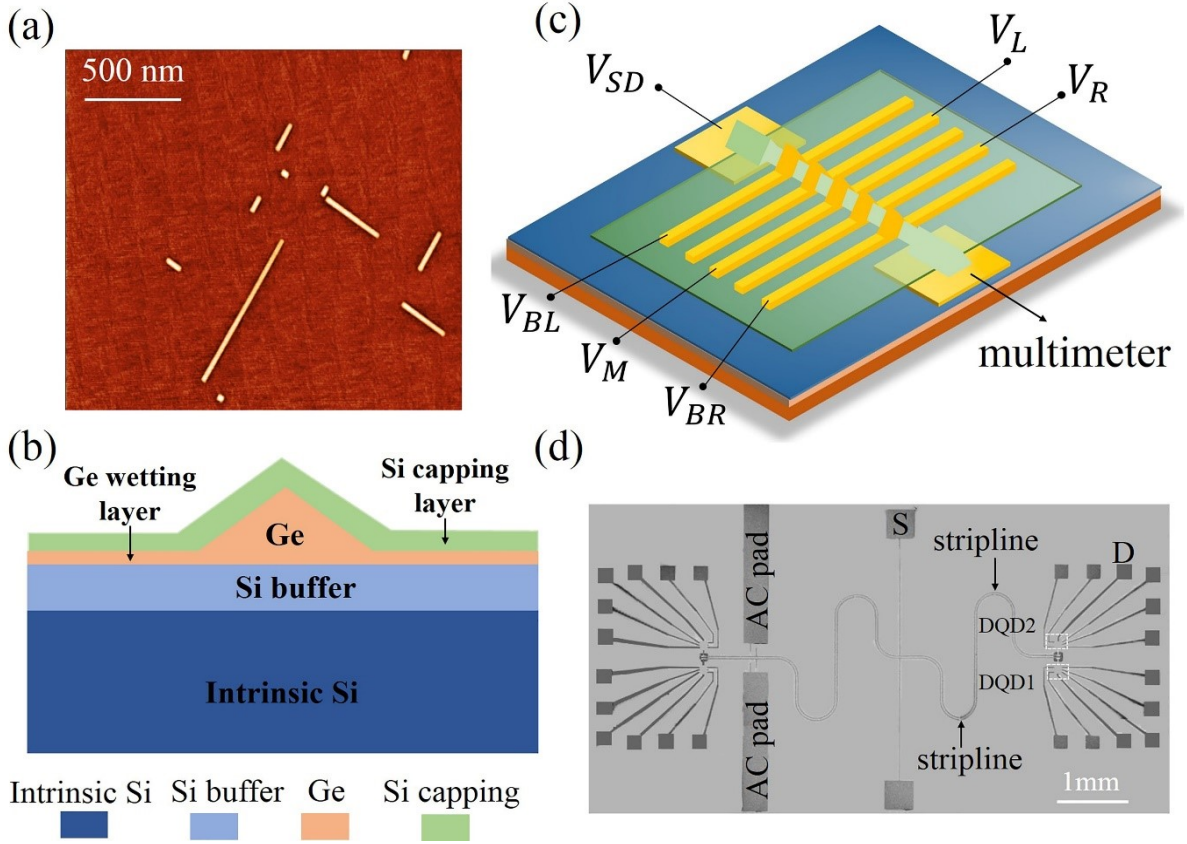

**Figure A.** (a) Atomic force microscopy image of Ge hut wires which that were epitaxially grown by means of the SK growth mechanism. (b) Longitudinal section of the device along the center of the Ge nanowire. The Ge nanowire is grown on an epitaxial Si buffer layer and protected by a thin Si capping layer. (c) 3D schematic representation of the DQD. (d) SEM image of our sample structure: half-wavelength reflection line resonator integrated with four DQDs; only the DQD2 is used in this study.

## Appendix B. Ge double quantum dot

In figure B, we show the stability diagram of a DQD coupled to the resonator by the source lead at $V_{SD}$=1.5 mV with barrier gate voltages $V_{BL}$=1 V, $V_M$=1.25 V and $V_{BR}$=1 V. A regular pattern of bias triangle pairs is clearly visible, from which we can extract the charging energies of the left dot $U_L$=4.5 meV and the right dot $U_R$=4.3 meV. The lever arms for conversion of gate voltages into energies are $\alpha_L$= 0.13 and $\alpha_R$= 0.14. The size and shape of the bias triangles are consistent over the whole range of the measurement. This result underlines the high degree of control over the electrochemical potentials of the QDs as well as the tunnel couplings.

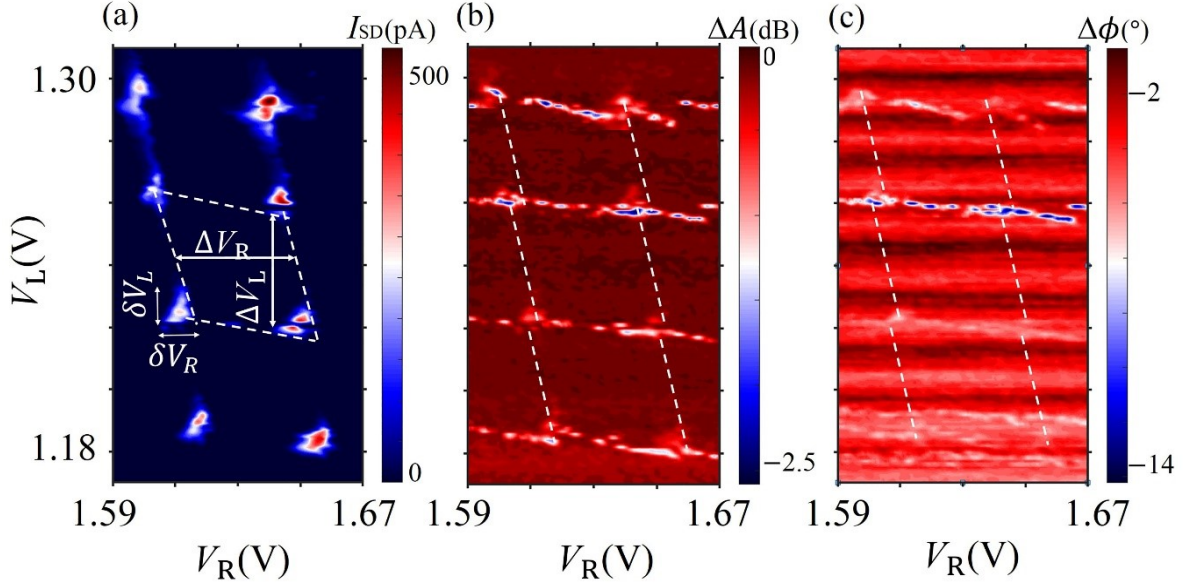

**Figure B.** (a) Stability transport signal diagram of the DQD. Hole tunneling between the left (right) dot and reservoir (drain) is indicated by the white dashed line. $\Delta V_L$ ($\Delta V_R$) denotes the variation in the plunger gate voltage required to populate or deplete the single hole in the left (right) dot. (b) Amplitude shift and (c) phase response as functions of $V_L$ and $V_R$ over the same gate voltage range as in (a).

## Appendix C. Modelling of the DQD-resonator interaction

The amplitude and phase signals from the resonator as functions of frequency were extracted using the network analyzer. When the DQD was tuned in the Coulomb blockade regime, a $\lambda/2$ open-circuit microstrip model [40, 42] was applied to describe the resonator. The reflection coefficient can be expressed as:

$$S_{11}(\omega) = -\frac{j(\omega_0-\omega)+g_{\text{eff}}\chi+\frac{1}{2}(\kappa_i-\kappa_e)}{j(\omega_0-\omega)+g_{\text{eff}}\chi+\frac{1}{2}(\kappa_i+\kappa_e)}. \tag{C1}$$

which determines the amplitude and phase, $A=|S_{11}|$ and $\phi=\arg(S_{11})$. From this model, we can obtain the resonance frequency $\omega_0/2\pi = 6.038$ GHz, the internal loss $\kappa_i/2\pi = 2.6$ MHz, the external loss $\kappa_e/2\pi = 4.0$ MHz, the total photon loss rate $\kappa/2\pi = \kappa_i/2\pi + \kappa_i/2\pi = 6.6$ MHz, and the quality factor $Q = \omega_0/\kappa$ of 820.

Considering the DQD energy scale, our system can be seen as a quantum two-level dipole coupled to a resonator, and it can be interpreted by the master equation based on the Jaynes-Cummings model

[45]. First, we write the total Hamiltonian of the system, which reads:

$$H = \hbar\Delta_0 a^+ a + \frac{\hbar\Delta}{2}\sigma_Z + \hbar g_{\text{eff}}(\sigma_+ a + \sigma_- a^+). \tag{C2}$$

Here, $\omega_0$ is the resonance frequency of the resonator, $\omega$ is the probe frequency, $g_{\text{eff}} = g_C \frac{2t_C}{\Omega}$ is the effective coupling strength between the DQD and the resonator. $\Omega = \sqrt{(2t_C)^2 + \varepsilon^2}$, $2t_C$ is the interdot tunneling rate of the DQD, $\Delta_0 = \omega_0 - \omega$, and $\Delta = \Omega - \omega_0$. For the whole system, the dissipation of energy mainly consists of the internal and external resonator dissipation rates $\kappa_i$ and $\kappa_e$ and the decoherence rate $\gamma$. We use a Markovian master equation approach to describe the dynamics of the system:

$$\dot{\rho} = -i[H,\rho] + \kappa D[a]\rho + \gamma D[\sigma_-]\rho \tag{C3}$$

$D[a]\rho = a\rho a^+ - \frac{1}{2}a^+ a\rho - \frac{1}{2}\rho a^+ a$, which is named the Lindblad operator in the literature, and the total loss rate $\kappa = \kappa_i + \kappa_e$. We obtain:

$$\dot{a} = -i\Delta_0 a - ig_{\text{eff}}\sigma_- - \frac{1}{2}\kappa a + \sqrt{\kappa_e}a_{\text{in}}, \tag{C4}$$

$$\dot{\sigma}_- = -i\Delta\sigma_- + ig_{\text{eff}}a\sigma_Z - \frac{1}{2}\gamma\sigma_-. \tag{C5}$$

In this measurement, we assume that the QD stays near its lower energy state with high probability, therefore, $\sigma_Z \to -1$. Taking a rotating wave approximation for $\dot{a} = -i(\omega_0 - \omega)a$, we find:

$$-i(\omega_0 - \omega)a = -i\Delta_0 a - ig_{\text{eff}}\sigma_- - \frac{1}{2}\kappa a + \sqrt{\kappa_e}a_{\text{in}}, \tag{C6}$$

$$-i(\omega_0 - \omega)\sigma_- = -i\Delta\sigma_- - ig_{\text{eff}}a - \frac{1}{2}\gamma\sigma_-. \tag{C7}$$

According to the input-output constraint condition $a_{\text{in}} + a_{\text{out}} = \sqrt{\kappa_e}a$ [34], we obtain:

$$S_{11}(\omega) = -\frac{j(\omega_0-\omega) + g_{\text{eff}}\chi + \frac{1}{2}(\kappa_i - \kappa_e)}{j(\omega_0-\omega) + g_{\text{eff}}\chi + \frac{1}{2}(\kappa_i + \kappa_e)}. \tag{C8}$$

Here $\chi = \frac{g_{\text{eff}}}{j(\Omega-\omega)+\gamma}$ characterizes the DQD susceptibility to the microwave wave field, and $g_{\text{eff}} = g_C\frac{2t_C}{\Omega}$. After obtaining the parameters $\kappa_i$ and $\kappa_e$, we can extract the remaining parameters $g_C$, $2t_C$ and $\gamma$ by further fitting $\Delta\phi$ as a function of the detuning $\varepsilon$. This fitting method has been used in previous experimental studies of graphene systems [34, 35], which have shown the accuracy.